\pdfoutput=1
\documentclass[letterpaper]{article}
\usepackage[preprint]{aaai2027}
\usepackage[hyphens]{url}
\usepackage{graphicx}
\usepackage{booktabs}
\usepackage{amsmath}
\usepackage{amssymb}
\usepackage{natbib}
\usepackage{caption}
\usepackage{xspace}
\xspaceaddexceptions{\%}
\newcommand{\GApis}{517\xspace}
\newcommand{\GMethods}{23,852\xspace}
\newcommand{\GScoped}{23,603\xspace}
\newcommand{\GMulti}{10,859\xspace}
\newcommand{\GMultiPct}{46.0\xspace}

\newcommand{\GMaxAlt}{73\xspace}

\newcommand{\GposetApis}{202\xspace}
\newcommand{\GposetClasses}{129\xspace}
\newcommand{\GposetPairs}{8,952\xspace}
\newcommand{\GposetInc}{2,077\xspace}
\newcommand{\GposetIncPct}{23.2\xspace}
\newcommand{\GposetEq}{5,879\xspace}
\newcommand{\GposetWide}{45\xspace}
\newcommand{\GposetWidePct}{34.9\xspace}
\newcommand{\GposetMaxWidth}{18\xspace}
\newcommand{\WGmail}{7\xspace}
\newcommand{\SGmail}{14\xspace}

\newcommand{\WCal}{7\xspace}

\newcommand{\WChat}{18\xspace}

\newcommand{\GtasksR}{2,822\xspace}

\newcommand{\GnonuniqRPct}{3.2\xspace}

\newcommand{\GnonuniqKtwo}{4.0\xspace}

\newcommand{\GnonuniqKthree}{3.7\xspace}

\newcommand{\GnonuniqKfive}{3.4\xspace}

\newcommand{\GnonuniqKLo}{3.4\xspace}
\newcommand{\GnonuniqKHi}{4.0\xspace}
\newcommand{\GtasksK}{8,340\xspace}
\newcommand{\GwfN}{23,448\xspace}

\newcommand{\GwfSuccLo}{98.8\xspace}
\newcommand{\GwfSuccHi}{99.5\xspace}
\newcommand{\GwfApis}{397\xspace}
\newcommand{\GwfWorstPct}{5.5\xspace}

\newcommand{\GscalPairs}{150\xspace}
\newcommand{\GscalConf}{43\xspace}
\newcommand{\GscalConfPct}{28.7\xspace}
\newcommand{\GscalTopA}{31\xspace}
\newcommand{\GscalTopB}{25\xspace}
\newcommand{\GscalTopC}{19\xspace}
\newcommand{\GscalSubLo}{0.7\xspace}
\newcommand{\GscalSubHi}{20.7\xspace}
\newcommand{\GscalSubNZ}{5\xspace}
\newcommand{\GscalSubTot}{6\xspace}
\newcommand{\GscalNoServices}{12.7\xspace}
\newcommand{\Gsixstrict}{98.7\xspace}
\newcommand{\GsixstrictD}{96.8\xspace}
\newcommand{\GsixstrictBase}{94.2\xspace}
\newcommand{\GsixstrictDiff}{+4.5\xspace}
\newcommand{\Gsixverb}{92.0\xspace}
\newcommand{\GsixverbD}{83.9\xspace}
\newcommand{\GsixverbBase}{85.9\xspace}
\newcommand{\GsixverbDiff}{+6.1\xspace}
\newcommand{\Gsixwriteheavy}{90.7\xspace}
\newcommand{\GsixwriteheavyD}{80.6\xspace}
\newcommand{\GsixwriteheavyBase}{78.6\xspace}
\newcommand{\GsixwriteheavyDiff}{+12.1\xspace}
\newcommand{\GsixPairs}{150\xspace}
\newcommand{\GsixDistinct}{62\xspace}
\newcommand{\GsixCmp}{1,181\xspace}
\newcommand{\GsixBaseLo}{78.6\xspace}
\newcommand{\GsixBaseHi}{94.2\xspace}
\newcommand{\MPerms}{919\xspace}
\newcommand{\MEndpoints}{22,305\xspace}

\newcommand{\MDWEp}{10,987\xspace}
\newcommand{\MDWInc}{99.7\xspace}
\newcommand{\MDWTasks}{493\xspace}
\newcommand{\MDWNonuniq}{29.8\xspace}
\newcommand{\MDWPairs}{565\xspace}
\newcommand{\MDWCollapsed}{67.4\xspace}
\newcommand{\MDWCollapsedD}{61.9\xspace}
\newcommand{\MDWBase}{61.0\xspace}
\newcommand{\MDWDiff}{+6.4\xspace}
\newcommand{\MDWCmpN}{100\xspace}
\newcommand{\MDWIncD}{378\xspace}

\newcommand{\MAppEp}{8,806\xspace}
\newcommand{\MAppInc}{99.7\xspace}
\newcommand{\MAppTasks}{450\xspace}
\newcommand{\MAppNonuniq}{30.7\xspace}
\newcommand{\MAppPairs}{363\xspace}
\newcommand{\MAppCollapsed}{74.4\xspace}
\newcommand{\MAppCollapsedD}{67.7\xspace}
\newcommand{\MAppBase}{61.1\xspace}
\newcommand{\MAppDiff}{+13.3\xspace}
\newcommand{\MAppCmpN}{72\xspace}
\newcommand{\MAppIncD}{198\xspace}

\newcommand{\MDPEp}{2,512\xspace}

\newcommand{\MDPTasks}{24\xspace}
\newcommand{\MDPNonuniq}{33.3\xspace}
\newcommand{\MDPPairs}{13\xspace}
\newcommand{\MDPCollapsed}{46.2\xspace}
\newcommand{\MDPCollapsedD}{37.5\xspace}
\newcommand{\MDPBase}{50.0\xspace}
\newcommand{\MDPDiff}{$-$3.8\xspace}
\newcommand{\MDPCmpN}{8\xspace}
\newcommand{\MDPIncD}{8\xspace}

\newcommand{\MDWAdmin}{148\xspace}

\newcommand{\MCollapsedLo}{46.2\xspace}
\newcommand{\MCollapsedHi}{74.4\xspace}
\newcommand{\MCollapsedBigLo}{67.4\xspace}
\newcommand{\MCollapsedBigHi}{74.4\xspace}

\newcommand{\MNonuniqLo}{29.8\xspace}
\newcommand{\MNonuniqHi}{33.3\xspace}
\newcommand{\GrMed}{2.0\xspace}

\newcommand{\GrGeTen}{16.7\xspace}

\newcommand{\GrN}{24\xspace}
\newcommand{\GrTaskN}{90\xspace}
\newcommand{\GrTaskMed}{49.4\xspace}
\newcommand{\MDWrMed}{3.6\xspace}
\newcommand{\MDWrMax}{44.2\xspace}

\newcommand{\MDWrGeTen}{27.5\xspace}

\newcommand{\MDWrN}{51\xspace}

\newcommand{\MDWrTaskMed}{3.6\xspace}
\newcommand{\MApprMed}{3.1\xspace}

\newcommand{\MApprN}{45\xspace}

\newcommand{\MDPrN}{3\xspace}

\newcommand{\SrcGN}{150\xspace}
\newcommand{\SrcGBe}{12.7\xspace}
\newcommand{\SrcGDc}{46.0\xspace}
\newcommand{\SrcGGr}{41.3\xspace}

\newcommand{\SrcMDWN}{565\xspace}
\newcommand{\SrcMDWBe}{6.2\xspace}
\newcommand{\SrcMDWDc}{33.1\xspace}
\newcommand{\SrcMDWGr}{60.7\xspace}

\newcommand{\SrcMAppN}{363\xspace}
\newcommand{\SrcMAppBe}{5.2\xspace}
\newcommand{\SrcMAppDc}{27.8\xspace}
\newcommand{\SrcMAppGr}{66.9\xspace}

\newcommand{\SrcBeLo}{5.2\xspace}
\newcommand{\SrcBeHi}{12.7\xspace}
\newcommand{\SrcDcLo}{27.8\xspace}
\newcommand{\SrcDcHi}{46.0\xspace}
\newcommand{\SrcGrLo}{41.3\xspace}
\newcommand{\SrcGrHi}{66.9\xspace}

\newcommand{\CIDWNonuniqLo}{25.9\xspace}
\newcommand{\CIDWNonuniqHi}{34.0\xspace}

\newcommand{\DiffSig}{4\xspace}
\newcommand{\DiffTot}{6\xspace}

\newcommand{\DiffSigLo}{4.5\xspace}
\newcommand{\DiffSigHi}{13.3\xspace}
\newcommand{\BaseAllLo}{50\xspace}
\newcommand{\BaseAllHi}{94\xspace}
\newcommand{\AgN}{1,176\xspace}
\newcommand{\AgModels}{6\xspace}
\newcommand{\AgAnti}{98\xspace}
\newcommand{\AgNmUnan}{39.8\xspace}
\newcommand{\AgNmPair}{66.5\xspace}
\newcommand{\AgNmNarrow}{54.6\xspace}
\newcommand{\AgNmWide}{20.6\xspace}
\newcommand{\AgNmGeTwo}{20.7\xspace}

\newcommand{\AgRcUnan}{43.9\xspace}

\newcommand{\AgRcNarrow}{79.9\xspace}

\newcommand{\AgRcPmLo}{53.1\xspace}
\newcommand{\AgRcPmHi}{100.0\xspace}
\newcommand{\AgMoved}{33.2\xspace}
\newcommand{\AgUnanGain}{4.1\xspace}
\newcommand{\AgNarrowGain}{25.3\xspace}

\newcommand{\MxTasks}{285\xspace}
\newcommand{\MxN}{1,708\xspace}
\newcommand{\MxMinimal}{99.0\xspace}
\newcommand{\MxNarrow}{74.1\xspace}
\newcommand{\MxGap}{24.9\xspace}
\newcommand{\MxGeTwo}{18.1\xspace}

\newcommand{\MxGapLo}{19.6\xspace}
\newcommand{\MxGapHi}{38.6\xspace}
\newcommand{\MxMinLo}{98.2\xspace}
\newcommand{\MxMinHi}{100.0\xspace}
\newcommand{\MxWorstMin}{99.3\xspace}
\newcommand{\MxWorstNarrow}{60.7\xspace}
\newcommand{\ScAll}{73.7\xspace}
\newcommand{\ScK}{5\xspace}
\newcommand{\ScN}{1,497\xspace}
\newcommand{\ScLo}{46.0\xspace}
\newcommand{\ScHi}{92.0\xspace}
\newcommand{\GdistKtwo}{71\xspace}
\newcommand{\GdistKthree}{66\xspace}
\newcommand{\GdistKfive}{60\xspace}

\newcommand{\GnonuniqTasksR}{90\xspace}
\newcommand{\GsixCmpD}{197\xspace}
\newcommand{\GsixstrictBaseD}{81.7\xspace}
\newcommand{\GsixverbBaseD}{71.1\xspace}
\newcommand{\GsixwriteheavyBaseD}{72.1\xspace}
\newcommand{\MDWBaseD}{62.5\xspace}
\newcommand{\MDWCmpD}{88\xspace}
\newcommand{\MDWNonuniqN}{147\xspace}
\newcommand{\MAppBaseD}{64.5\xspace}
\newcommand{\MAppCmpD}{62\xspace}

\newcommand{\MDPBaseD}{50.0\xspace}
\newcommand{\MDPCmpD}{2\xspace}

\newcommand{\CsDWStrict}{95.4\xspace}
\newcommand{\CsDWStrictBase}{66.0\xspace}
\newcommand{\CsDWStrictD}{93.1\xspace}
\newcommand{\CsDWStrictBaseD}{62.5\xspace}
\newcommand{\CsDWStrictDiff}{+29.4\xspace}
\newcommand{\CsDWVerb}{92.0\xspace}
\newcommand{\CsDWVerbBase}{66.0\xspace}
\newcommand{\CsDWVerbD}{88.1\xspace}
\newcommand{\CsDWVerbBaseD}{62.5\xspace}
\newcommand{\CsDWVerbDiff}{+26.0\xspace}
\newcommand{\CsDWWh}{77.7\xspace}
\newcommand{\CsDWWhBase}{57.0\xspace}
\newcommand{\CsDWWhD}{85.7\xspace}
\newcommand{\CsDWWhBaseD}{58.0\xspace}
\newcommand{\CsDWWhDiff}{+20.7\xspace}
\newcommand{\CsAppStrict}{93.7\xspace}
\newcommand{\CsAppStrictBase}{70.8\xspace}
\newcommand{\CsAppStrictD}{88.9\xspace}
\newcommand{\CsAppStrictBaseD}{67.7\xspace}
\newcommand{\CsAppStrictDiff}{+22.8\xspace}
\newcommand{\CsAppVerb}{89.0\xspace}
\newcommand{\CsAppVerbBase}{69.4\xspace}
\newcommand{\CsAppVerbD}{79.8\xspace}
\newcommand{\CsAppVerbBaseD}{67.7\xspace}
\newcommand{\CsAppVerbDiff}{+19.5\xspace}
\newcommand{\CsAppWh}{66.9\xspace}
\newcommand{\CsAppWhBase}{66.7\xspace}
\newcommand{\CsAppWhD}{74.7\xspace}
\newcommand{\CsAppWhBaseD}{64.5\xspace}
\newcommand{\CsAppWhDiff}{+0.3\xspace}

\newcommand{\CsGraphVerbLo}{89.0\xspace}
\newcommand{\CsGraphVerbHi}{92.0\xspace}
\newcommand{\CsGraphBaseLo}{57.0\xspace}
\newcommand{\CsGraphBaseHi}{70.8\xspace}
\newcommand{\CsAbove}{5\xspace}
\newcommand{\CsCells}{6\xspace}
\newcommand{\CsGoogleLo}{90.7\xspace}
\newcommand{\CsGoogleHi}{98.7\xspace}
\newcommand{\GscalConfD}{22.6\xspace}
\newcommand{\GscalDistinct}{62\xspace}
\newcommand{\MsDWN}{565\xspace}
\newcommand{\MsDWConf}{23.4\xspace}

\newcommand{\MsDWConfD}{30.7\xspace}
\newcommand{\MsDWDN}{378\xspace}
\newcommand{\MsDWBW}{8.1\xspace}
\newcommand{\MsDWUnsep}{1\xspace}
\newcommand{\MsAppN}{363\xspace}
\newcommand{\MsAppConf}{15.7\xspace}

\newcommand{\MsAppConfD}{23.7\xspace}
\newcommand{\MsAppDN}{198\xspace}
\newcommand{\MsAppBW}{5.2\xspace}

\newcommand{\MsDPN}{13\xspace}

\newcommand{\MsRobLo}{14.3\xspace}
\newcommand{\MsRobHi}{30.8\xspace}
\newcommand{\MsExBreadthA}{243\xspace}
\newcommand{\MsExWriteA}{153\xspace}
\newcommand{\MsExBreadthB}{297\xspace}
\newcommand{\MsExWriteB}{20\xspace}
\newcommand{\GscalConfRaw}{29.3\xspace}
\newcommand{\MsConfLo}{15.7\xspace}
\newcommand{\MsConfHi}{23.4\xspace}
\newcommand{\ExAnRoBreadth}{73\xspace}
\newcommand{\ExAnRoSvc}{3\xspace}
\newcommand{\ExAnEdBreadth}{148\xspace}
\newcommand{\ExAnEdSvc}{2\xspace}
\newcommand{\GsixstrictDiffD}{+15.0\xspace}
\newcommand{\GsixverbDiffD}{+12.8\xspace}
\newcommand{\GsixwriteheavyDiffD}{+8.6\xspace}
\newcommand{\MDWDiffD}{$-$0.6\xspace}
\newcommand{\MAppDiffD}{+3.2\xspace}
\newcommand{\MDPDiffD}{$-$12.5\xspace}
\newcommand{\CsGapMax}{5.0\xspace}
\newcommand{\CsGapMaxD}{7.9\xspace}
\newcommand{\MDWAdminSame}{90\xspace}
\newcommand{\ScCells}{300\xspace}
\newcommand{\ExBqBreadth}{286\xspace}
\newcommand{\ExBqWrite}{144\xspace}
\newcommand{\ExBqSvc}{10\xspace}
\newcommand{\ExCpBreadth}{565\xspace}
\newcommand{\ExCpWrite}{61\xspace}
\newcommand{\ExCpSvc}{27\xspace}
\newcommand{\CsDWStrictDiffD}{+30.6\xspace}
\newcommand{\CsDWVerbDiffD}{+25.6\xspace}
\newcommand{\CsDWWhDiffD}{+27.8\xspace}
\newcommand{\CsAppStrictDiffD}{+21.1\xspace}
\newcommand{\CsAppVerbDiffD}{+12.1\xspace}
\newcommand{\CsAppWhDiffD}{+10.2\xspace}
\newcommand{\ConfAllLo}{15.7\xspace}
\newcommand{\ConfAllHi}{28.7\xspace}
\newcommand{\AgRcTopN}{3\xspace}
\newcommand{\AgRcTopLo}{95.9\xspace}
\newcommand{\AgRcRestN}{3\xspace}
\newcommand{\AgRcRestHi}{76.5\xspace}
\newcommand{\FmMismatch}{0\xspace}
\newcommand{\SrcGND}{62\xspace}
\newcommand{\SrcGBeD}{16.1\xspace}
\newcommand{\SrcGDcD}{33.9\xspace}
\newcommand{\SrcGGrD}{50.0\xspace}

\newcommand{\SrcMDWND}{378\xspace}
\newcommand{\SrcMDWBeD}{7.1\xspace}
\newcommand{\SrcMDWDcD}{42.9\xspace}
\newcommand{\SrcMDWGrD}{50.0\xspace}

\newcommand{\SrcMAppND}{198\xspace}
\newcommand{\SrcMAppBeD}{7.6\xspace}
\newcommand{\SrcMAppDcD}{34.8\xspace}
\newcommand{\SrcMAppGrD}{57.6\xspace}

\newcommand{\SrcShiftMax}{12.1\xspace}
\newcommand{\FmNmBreadth}{54.6\xspace}
\newcommand{\FmNmWrite}{55.8\xspace}
\newcommand{\FmNmSvc}{67.5\xspace}
\newcommand{\FmNmAdmin}{93.9\xspace}
\newcommand{\FmNmAllFour}{38.9\xspace}

\newcommand{\FmNmAllFourEx}{63.3\xspace}
\newcommand{\FmNmAllFourGivenEx}{61.6\xspace}
\newcommand{\FmNmDom}{32.3\xspace}
\newcommand{\FmRcBreadth}{79.9\xspace}
\newcommand{\FmRcWrite}{75.5\xspace}
\newcommand{\FmRcSvc}{74.0\xspace}
\newcommand{\FmRcAdmin}{96.8\xspace}
\newcommand{\FmRcAllFour}{52.9\xspace}

\newcommand{\FmRcDom}{12.9\xspace}
\newcommand{\FmMxBreadth}{74.1\xspace}
\newcommand{\FmMxWrite}{73.1\xspace}
\newcommand{\FmMxSvc}{75.1\xspace}
\newcommand{\FmMxAdmin}{98.4\xspace}
\newcommand{\FmMxAllFour}{66.7\xspace}

\newcommand{\FmMxDom}{22.1\xspace}
\newcommand{\FmMxGapAllFour}{32.3\xspace}
\newcommand{\FmChecked}{1,878\xspace}
\newcommand{\FmNmN}{588\xspace}
\newcommand{\FmRcN}{588\xspace}
\newcommand{\FmMxN}{1,708\xspace}
\newcommand{\FmMxGapBreadth}{24.9\xspace}

\title{Incomparable Minima in Agent Tool Authorization}
\author{Xin Xu}
\affiliations{Carnegie Mellon University\\
xuxin@cmu.edu}

\begin{document}
\maketitle

\begin{abstract}
An agent that needs a tool needs authority to call it, and every proposal for
governing that authority has to answer one question: of two grants that both
let the agent finish a task, which one is smaller? Over-privilege benchmarks score agents against this answer, and permission solvers compute it. We check the question against the authorisation tables that
Microsoft and Google publish, covering \MEndpoints Microsoft Graph endpoint
bindings and \GMethods Google API methods, on tasks generated from each
table's own resource structure. For \MNonuniqLo--\MNonuniqHi\% of Graph tasks
and \GnonuniqRPct\% of Google resource tasks the question has no answer,
because the minimal sufficient grants form an antichain, a set of grants none
of which authorises a subset of what another does. In its two main authorisation
modes, Microsoft's shipped \texttt{privilegeLevel} rates
\MCollapsedBigLo--\MCollapsedBigHi\% of these incomparable pairs equal. Google ships no such scale, so on Google we apply a
reconstruction of the FORTIS privilege ladder, which rates
\CsGoogleLo--\CsGoogleHi\% equal. Applying the same reconstruction to Graph attributes much of that difference to the scale rather than the corpus: its form based on the HTTP verb alone rates \Gsixverb\% of Google pairs and
\CsGraphVerbLo--\CsGraphVerbHi\% of Graph pairs equal. Four risk measures (operations, writes, services reached, administrative surface) order \GscalConfPct\% of incomparable Google pairs and
\MsConfLo--\MsConfHi\% of Graph pairs in opposite directions, so for those pairs the ranking has no measure-independent answer. Offered pools of minimal and larger grants,
\AgModels frontier models pick a minimal grant on \MxMinimal\% of \MxN
decisions, a grant smallest under all four measures on \FmMxAllFour\%, and on
\FmMxDom\% a grant that another option matches or beats on every measure. We propose that benchmarks report which minimal grant an agent chose and that deployments show the alternatives to whoever approves the grant.
\end{abstract}

\section{Introduction}

Deployed agents call tools on a user's behalf, and the authority they hold
while doing so is now a governed object. Enterprise stacks scope OAuth tokens
per tool. The Model Context Protocol recommends a progressive scope model,
starting minimal and requesting more when needed, and requires that servers
``account for scope hierarchies, where a broader scope implies narrower ones''
when deciding whether a token suffices \cite{mcpauth}. Implication between
scopes is the structure that rule reasons over; this paper measures how much of
a real scope catalogue that structure covers. Recent work measures whether agents respect this discipline: benchmarks score an agent on whether it
reached for a higher-privilege tool when a lower-privilege one would have done
\cite{toolprivbench,fortis,authbench}, and frameworks compute a minimal
permission set for a plan and ask the user to approve it \cite{miniscope,
skillscope}.

Each of these systems has to decide, for a task and two sufficient grants, which grant is smaller, because the answer is what a score is computed against or what a solver returns. Each fixes that answer when building its inventory rather than measuring it on a real catalogue.

The constructions are stated plainly by their authors. FORTIS defines the
ground-truth skill as $s^\star = \arg\min_{s} \ell_S(s)$ over an ordinal scale
$\ell_S \in \{0,\dots,4\}$, and reports that ``the benchmark construction
ensur[es] that the minimum is unique at the label level'', a uniqueness
``ensured primarily through structural design rather than post-hoc
annotation'' \cite{fortis}. ToolPrivBench builds scenarios from a five-level
privilege schema and, during human review, discards cases in which ``the
privilege contrast is ambiguous'' \cite{toolprivbench}. MiniScope, which does
use a genuine partial order derived from OAuth scope containment, resolves the
remaining freedom with an integer program over a cost model and does not
discuss whether the optimum is unique \cite{miniscope}.

These are reasonable decisions for building a benchmark. Our question is what
they cost on real authorisation systems, where nobody has arranged for the
minimum to be unique. We answer it with the authorisation tables the vendors themselves publish, so
every table measurement below is a deterministic function of two public
downloads, with no model in the loop. The Google API Discovery Service states,
for every method of \GApis APIs, the set of OAuth scopes that authorise it, and
the semantics of that set is disjunctive: a method carrying several scopes
``supports one of'' them, and ``the decision of which scope to use is left to
the developer'' \cite{googlediscoveryusing}. Microsoft publishes the analogous
table for Graph, and additionally assigns almost every permission an ordinal
\texttt{privilegeLevel} and an administrator-consent flag \cite{msgraphperms}.

The two corpora therefore arrive with different scales. On Graph we test the
ordinal scale its vendor designed. Google publishes none, so on Google we test
a mechanical reconstruction of the FORTIS ladder; to keep the two from being
confounded, \S\ref{sec:scales} also applies that reconstruction to Graph, which
separates the effect of the scale from the effect of the corpus. In both
corpora the tasks are generated from the tables' own resource structure rather
than observed from deployed agents (\S\ref{sec:method}), so every rate we
report describes the published tables under those task families.

\paragraph{Contributions.}
\begin{itemize}\itemsep1pt
\item How often a task's minimal sufficient authority is an antichain rather
      than a point, with the order derived from the vendors' own tables, and
      how far apart the competing minima are in reach (\S\ref{sec:antichain},
      \S\ref{sec:perfect}).
\item A test of Microsoft's own scale and a reconstructed FORTIS ladder against
      a null baseline of pairs the scale can order correctly, with one ladder
      applied to both corpora to separate scale from corpus
      (\S\ref{sec:scales}).
\item Measure dependence on both corpora: two of four risk measures reverse
      the ranking on \GscalConfPct\% of Google pairs and
      \MsConfLo--\MsConfHi\% of Graph pairs (\S\ref{sec:measures}).
\item \AgN decisions by \AgModels frontier models choosing among enumerated grants, and \MxN more in pools mixing minimal and larger grants, scored under all four measures; in the mixed pools \FmMxDom\% of picks are Pareto-dominated (\S\ref{sec:format}).
\item A localisation to the single step: per-step acquisition reaches the
      whole-workflow minimum on \GwfSuccLo--\GwfSuccHi\% of \GwfN Google
      workflows (\S\ref{sec:composition}).
\end{itemize}

\begin{figure}[t]\centering
\includegraphics{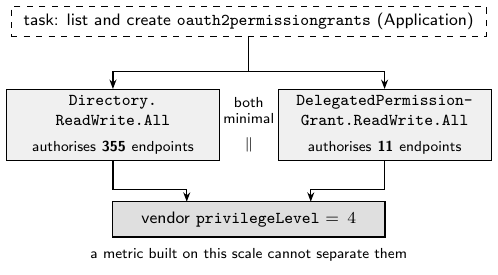}
\caption{One Microsoft Graph task. Two permissions are minimal and
sufficient, neither authorises a superset of the other, and Microsoft's
\texttt{privilegeLevel} assigns both the same value.}
\label{fig:collapse}
\end{figure}

\section{Ordinal Privilege Scales in Agent Evaluation}
\label{sec:background}

The systems we are testing share a shape. A task admits a set of sufficient
authorities; the system must pick one, or score whether the agent picked well;
and the picking is done by minimising a scalar.

FORTIS places skills and tools on the ladder \textsf{Observe},
\textsf{Read}, \textsf{Annotate/Modify}, \textsf{Execute},
\textsf{Bulk/Admin} and aggregates a set by its maximum, scoring an agent by
whether $\ell_S(\hat s) > \ell_S(s^\star)$ \cite{fortis}. ToolPrivBench
reports \textsf{OPUR@k}, the rate at which an agent uses a higher-privilege
tool while lower-privilege sufficient alternatives remain \cite{toolprivbench}.
AuthBench defines a genuine partial order, comparing policies on read, write
and execute axes, and then scalarises it, taking $\arg\min_\pi |\pi|$ \cite{authbench}. It does
note that the boundary ``may be non-unique'', attributing that to different
executors solving a task by different routes, which leaves the case we
measure, one task and one route, untouched.

Two features of this design matter for what follows: the scale is
\emph{total}, returning a comparison for every pair whether or not one exists,
and aggregation is by maximum, so two grants agreeing on their most privileged
member are indistinguishable. Neither is a defect where these systems were
built. FORTIS constructs its hierarchy so that ``only one skill occupies that
level for the relevant functional category'' \cite{fortis}, and within such an
inventory a total scale loses nothing.

FORTIS's validation data show the same mechanism at work. Its annotators agreed on 193 of
200 sampled labels, and it reports that the seven disagreements ``arose
primarily from borderline tool granularity, where reviewers differed on whether
a single coarse-grained tool versus multiple fine-grained tools constituted the
true minimum'' \cite{fortis}. Coarse against fine is the mechanism we isolate in
\S\ref{sec:sources} and measure at \SrcGrLo--\SrcGrHi\% of incomparable pairs;
where the construction did not fully suppress it, human experts disagreed on
which grant was smaller.

\section{Corpora and Method}
\label{sec:method}

\paragraph{Google API Discovery.}
Every Google API publishes a machine-readable discovery document listing its
OAuth2 scopes and, per method, the scopes that authorise it. We fetched all
available documents, giving \GApis APIs and \GMethods methods, of which
\GScoped carry at least one scope. Of those, \GMulti (\GMultiPct\%) list more
than one authorising scope, with up to \GMaxAlt alternatives for a single
method. Google's semantics for that list is disjunctive, so each method comes
with its own set of alternative authorisations, published by the vendor.

\paragraph{Microsoft Graph.}
Microsoft publishes a single permissions file binding \MPerms named
permissions to endpoints, keyed by authorisation scheme. We treat
\textsf{DelegatedWork}, \textsf{Application} and \textsf{DelegatedPersonal}
separately, since a grant in one is not a grant in another, giving
\MDWEp, \MAppEp and \MDPEp endpoint bindings. Each binding also carries a
\texttt{least=} annotation naming the schemes for which that permission is
Microsoft's own least-privilege recommendation; we do not use it, since the
object here is the order the table induces rather than the vendor's
recommendation over it. The file records no conjunctive dependencies between
permissions, so each is treated as an alternative in its own right. Almost
every permission carries a vendor-assigned \texttt{privilegeLevel} in
$\{1,\dots,4\}$ and a \texttt{requiresAdminConsent} flag.
\citet{lostinpermissions} measure the consent grants real M365 apps request; we
measure the authorisation table those grants are drawn from.

\paragraph{The order.}
We assign no privilege levels. For a scope or permission $s$, let $M(s)$ be the
set of operations the vendor says $s$ authorises, and order by coverage:
\[ s \le t \iff M(s) \subseteq M(t). \]
The order is computed from the vendor's table alone.
MiniScope builds the same containment relation over the scopes of ten
applications to solve over it \cite{miniscope}; we measure it across \GApis
APIs and \MPerms permissions. Scopes with identical coverage are equivalent
rather than incomparable, so widths are computed on the quotient.

\paragraph{Grants, sufficiency, minimality.}
A task $T$ is a set of operations. A grant is a set $F$ of scopes, and its
\emph{authority footprint} is $A(F) = \bigcup_{s \in F} M(s)$, the operations
$F$ lets its holder call. $F$ is sufficient for $T$ iff $T \subseteq A(F)$. We
order grants by footprint, $F \sqsubseteq G \iff A(F) \subseteq A(G)$, because
$A(F)$ is the blast radius of holding $F$ and is what a security argument is
about. Under this order sufficiency is upward closed, so the candidates for
least authority are the minimal sufficient grants. An \emph{antichain} is a set
of grants no two of which are ordered, and a task's minimal sufficient grants
always form one; the task has a unique minimum exactly when that antichain has
one member. We enumerate the minimal sufficient grants exactly, as minimal
hitting sets of the per-operation alternative sets. No task reached
the search's node budget. Where several grants share one footprint we keep the
smallest, breaking ties lexicographically, so every result is reproducible run
to run. Ordering by footprint rather than by name matters: by name inclusion
$\{\texttt{adsense}\}$ and $\{\texttt{adsense.readonly}\}$ would count as
incomparable, whereas by footprint they are comparable.

\paragraph{Privilege scales.}
Section~\ref{sec:scales} applies ordinal scales to grants, and the two corpora
do not come with the same kind of scale. On Graph we use the vendor's
\texttt{privilegeLevel} unchanged, aggregated over a grant by maximum as
FORTIS aggregates. Google publishes no level, so we reconstruct the FORTIS
ladder, levels 0 to 4, mechanically from each operation's HTTP verb and name,
in three variants. \emph{Strict} assigns level 4 to names matching
administrative or bulk patterns and to \texttt{DELETE}, then 3 to
\texttt{POST}, 2 to \texttt{PATCH} and \texttt{PUT}, and 0 or 1 to reads
depending on whether they list or fetch. \emph{Verb-only} reads the HTTP verb
alone. \emph{Write-heavy} assigns 4 to administrative names and 3 to every
state-changing verb. The verb-only variant uses nothing Graph does not publish
in the same form, so it is the same function on both corpora; for the other two
the name is the endpoint path on Graph and a list is a \texttt{GET} on a
collection.

\paragraph{Task generators.}
Tasks are generated from the tables rather than observed from agents. Google
resource tasks are the full method set of each REST resource with between 2
and 25 scoped methods; Google random tasks draw 20 samples of size $k$ from each
API with at least five scoped methods; Graph tasks group endpoints by their
first two path components and keep groups of 2 to 12. Each family encodes a
plausible unit of work, and the rates below are rates over these families: they
say how often a task built from a vendor's own resource grouping has no unique
minimum, not how often a deployed agent meets one, which depends on a workload
distribution no public trace records.

\paragraph{Units of analysis.}
Three units appear below and each table names its own. Non-uniqueness
(Table~\ref{tab:antichain}) is a rate over tasks, the unit an agent faces. Pair
statistics (Tables~\ref{tab:scales} and~\ref{tab:sources}, \S\ref{sec:measures})
pool the pairs of incomparable minima of every task, so a pair that is minimal
for several nested resources counts once per task; each is also reported per
distinct pair, counting each pair of footprints once. Reach ratios
(\S\ref{sec:perfect}) are computed per distinct antichain, since one antichain
repeated across many tasks would dominate a median. The comparable baseline of
\S\ref{sec:scales} comes from the same tasks and enumeration: it pairs each
minimal sufficient grant with every sufficient candidate the hitting-set search
returned whose footprint strictly contains it. Those candidates cannot drop a
scope and remain sufficient, so they are the non-minimal grants an agent would
plausibly request. Every count below is produced by a script from the two
public downloads; only \S\ref{sec:format} calls models.

\section{Incomparability in Published Tables}
\label{sec:incomparability}

On Google, restricting to the \GposetApis APIs with at least two scopes and
quotienting by identical coverage leaves \GposetClasses APIs with a
non-trivial order. Of \GposetPairs scope pairs, \GposetInc (\GposetIncPct\%)
are incomparable and \GposetEq have identical coverage. The quotient order has
width at least two for \GposetWide of the \GposetClasses APIs
(\GposetWidePct\%), reaching width \GposetMaxWidth. The services an agent
actually touches are among the widest: Gmail has width \WGmail over \SGmail
scopes, Calendar \WCal, and Google Chat \WChat, which separates message,
space and membership access.

On Graph the catalogue is partitioned by functional area, so \MDWInc\% of
permission pairs are incomparable under \textsf{DelegatedWork} and \MAppInc\%
under \textsf{Application}, mostly because unrelated areas do not overlap. The relevant question is therefore whether the permissions competing
to authorise \emph{the same task} are comparable.

\section{Minimal Sufficient Authority for a Task}
\label{sec:antichain}

\begin{table}[t]\centering\small
\begin{tabular}{@{}lrrr@{}}
\toprule
Corpus and task family & tasks & non-uniq. & antichains \\
\midrule
Graph, DelegatedWork     & \MDWTasks  & \MDWNonuniq\%  & \MDWrN \\
Graph, Application       & \MAppTasks & \MAppNonuniq\% & \MApprN \\
Graph, DelegatedPersonal & \MDPTasks  & \MDPNonuniq\%  & \MDPrN \\
\midrule
Google, resource sets   & \GtasksR  & \GnonuniqRPct\%  & \GrN \\
Google, random 2-method & \GtasksK  & \GnonuniqKtwo\%  & \GdistKtwo \\
Google, random 3-method & \GtasksK  & \GnonuniqKthree\%& \GdistKthree \\
Google, random 5-method & \GtasksK  & \GnonuniqKfive\% & \GdistKfive \\
\bottomrule
\end{tabular}
\caption{How often a task's minimal sufficient authority is an antichain
rather than a point. \emph{Non-uniq.}: share of tasks with two or more
incomparable minimal sufficient grants. \emph{Antichains}: number of distinct
antichains behind those tasks, since one antichain can be the minimum for
several tasks; for the Graph and Google resource rows it is the denominator of
the reach ratios in \S\ref{sec:perfect}. Tasks are exact enumerations.}
\label{tab:antichain}
\end{table}

On Microsoft Graph, \MNonuniqLo--\MNonuniqHi\% of tasks have no unique
minimal sufficient authority, depending on the authorisation mode
(Table~\ref{tab:antichain}). On Google the rate is an order of magnitude lower,
\GnonuniqRPct\% for resource tasks and \GnonuniqKLo--\GnonuniqKHi\% for random
$k$-method tasks, consistent with Google's finer-grained per-service scopes.
The non-unique tasks resolve to fewer distinct antichains, \GrN behind \GnonuniqTasksR Google resource tasks and \MDWrN behind
\MDWNonuniqN Graph \textsf{DelegatedWork} tasks, since one antichain is
often the minimum for a resource and its nested sub-resources, or for several
path groups covered by the same permissions.

A choice among incomparable minima always changes what the agent holds, and
this follows from the definition rather than from the data: minimal footprints
are mutually non-containing and all contain the task, so two minima that
authorised the same operations beyond the task would be the same grant. What
the data has to supply is how far apart they are, which is
\S\ref{sec:perfect}, and what makes them differ, which is \S\ref{sec:sources}.

Two Google examples make the shape concrete. For the task of reading BigQuery
project metadata, two grants are minimal and incomparable: \texttt{bigquery},
which authorises \ExBqBreadth methods across \ExBqSvc services, and \texttt{cloud-platform.read-only}, which authorises \ExCpBreadth methods across \ExCpSvc. The
first is narrower but carries BigQuery's write surface; the second is wider but
is the vendor's read-only credential. Which is less privileged depends on which
risk the deployment cares about, and the two answers diverge. Listing Google
Analytics account summaries, a single method, is minimally authorised by
\texttt{analytics.readonly} (\ExAnRoBreadth methods, \ExAnRoSvc services) or
by \texttt{analytics.edit} (\ExAnEdBreadth methods, \ExAnEdSvc services), and
again neither dominates.

\section{Ordinal Scales on Incomparable Grants}
\label{sec:scales}

\begin{table*}[t]\centering\footnotesize
\setlength{\tabcolsep}{4.5pt}
\begin{tabular}{@{}llrrrrrrrr@{}}
\toprule
 & & \multicolumn{4}{c}{per pair, pooled over tasks} & \multicolumn{4}{c}{per distinct pair} \\
\cmidrule(lr){3-6}\cmidrule(l){7-10}
Corpus and mode & Scale & incomp. & comparable & diff. & $n$ & incomp. & comparable & diff. & $n$ \\
\midrule
Graph, DelegatedWork     & vendor level & \MDWCollapsed\%  & \MDWBase\%  & \MDWDiff  & \MDWPairs/\MDWCmpN  & \MDWCollapsedD\%  & \MDWBaseD\%  & \MDWDiffD  & \MDWIncD/\MDWCmpD \\
Graph, Application       & vendor level & \MAppCollapsed\% & \MAppBase\% & \MAppDiff & \MAppPairs/\MAppCmpN & \MAppCollapsedD\% & \MAppBaseD\% & \MAppDiffD & \MAppIncD/\MAppCmpD \\
Graph, DelegatedPersonal & vendor level & \MDPCollapsed\%  & \MDPBase\%  & \MDPDiff  & \MDPPairs/\MDPCmpN  & \MDPCollapsedD\%  & \MDPBaseD\%  & \MDPDiffD  & \MDPIncD/\MDPCmpD \\
\midrule
Graph, DelegatedWork & ladder, strict      & \CsDWStrict\% & \CsDWStrictBase\% & \CsDWStrictDiff & \MDWPairs/\MDWCmpN & \CsDWStrictD\% & \CsDWStrictBaseD\% & \CsDWStrictDiffD & \MDWIncD/\MDWCmpD \\
                     & ladder, verb-only   & \CsDWVerb\%   & \CsDWVerbBase\%   & \CsDWVerbDiff   & & \CsDWVerbD\%   & \CsDWVerbBaseD\%   & \CsDWVerbDiffD & \\
                     & ladder, write-heavy & \CsDWWh\%     & \CsDWWhBase\%     & \CsDWWhDiff     & & \CsDWWhD\%     & \CsDWWhBaseD\%     & \CsDWWhDiffD & \\
Graph, Application   & ladder, strict      & \CsAppStrict\% & \CsAppStrictBase\% & \CsAppStrictDiff & \MAppPairs/\MAppCmpN & \CsAppStrictD\% & \CsAppStrictBaseD\% & \CsAppStrictDiffD & \MAppIncD/\MAppCmpD \\
                     & ladder, verb-only   & \CsAppVerb\%   & \CsAppVerbBase\%   & \CsAppVerbDiff   & & \CsAppVerbD\%   & \CsAppVerbBaseD\%   & \CsAppVerbDiffD & \\
                     & ladder, write-heavy & \CsAppWh\%     & \CsAppWhBase\%     & \CsAppWhDiff     & & \CsAppWhD\%     & \CsAppWhBaseD\%     & \CsAppWhDiffD & \\
\midrule
Google, resource sets & ladder, strict      & \Gsixstrict\%     & \GsixstrictBase\%     & \GsixstrictDiff     & \GsixPairs/\GsixCmp & \GsixstrictD\%     & \GsixstrictBaseD\%     & \GsixstrictDiffD     & \GsixDistinct/\GsixCmpD \\
                      & ladder, verb-only   & \Gsixverb\%       & \GsixverbBase\%       & \GsixverbDiff       & & \GsixverbD\%       & \GsixverbBaseD\%       & \GsixverbDiffD       & \\
                      & ladder, write-heavy & \Gsixwriteheavy\% & \GsixwriteheavyBase\% & \GsixwriteheavyDiff & & \GsixwriteheavyD\% & \GsixwriteheavyBaseD\% & \GsixwriteheavyDiffD & \\
\bottomrule
\end{tabular}
\caption{Share of grant pairs a scale rates equal. \emph{Incomp.}: pairs of
incomparable minimal sufficient grants, where no correct order exists.
\emph{Comparable}: the null baseline, pairs from the same tasks in which one
sufficient grant's footprint strictly contains the other's, so a correct order
exists (\S\ref{sec:method}). \emph{Diff.}: percentage points. $n$: incomparable
/ comparable pairs. The vendor rows audit Microsoft's own scale; the ladder rows
apply our reconstruction of the FORTIS ladder, to Graph on the same pairs as
the vendor rows and to Google, which ships no scale. The
\textsf{DelegatedPersonal} mode has too few pairs for the ladder rows.}
\label{tab:scales}
\end{table*}

A total scale must return something for an incomparable pair. It can rank the
pair, in which case the rank is the scale's own tie-break, or it can give both
grants the same level, which we call \emph{collapse}, in which case the choice
is invisible to any metric built on the scale. Table~\ref{tab:scales} counts
collapse on incomparable minima and on the comparable baseline of
\S\ref{sec:method}, which separates incomparability from general coarseness.

\paragraph{The vendor scale on Graph.}
The top block uses Microsoft's own \texttt{privilegeLevel}, applied to the pairs its own tables
generate and aggregated by maximum as agent benchmarks aggregate. It rates
\MDWCollapsed\% and \MAppCollapsed\% of incomparable minima equal in the two modes with hundreds of pairs. On comparable pairs from the same tasks, where the question
has an answer, it rates \MDWBase\% and \MAppBase\% equal. A two-proportion test separates this excess over the baseline from zero in \DiffSig of the \DiffTot vendor and Google
ladder strata, and where it separates it is worth \DiffSigLo--\DiffSigHi\
percentage points against baselines of \BaseAllLo--\BaseAllHi\%. Pairs are
clustered by task, so these are nominal tests; counted once per distinct pair,
the vendor scale's excess on Graph is \MDWDiffD and \MAppDiffD points.
The level already ties most comparable pairs, and ties incomparable minima at a similar or higher rate.

\paragraph{Scale versus corpus.}
The Google rows sit higher, at \CsGoogleLo--\CsGoogleHi\%, either because
Google's antichains are harder to separate or because the reconstructed ladder
resolves less than Microsoft's level. The middle block of
Table~\ref{tab:scales} decides between the two by applying the ladder to the
Graph pairs the vendor level was scored on. Under the verb-only
variant, the same function on both corpora, Graph's collapse rate rises to
\CsDWVerb\% and \CsAppVerb\%, against \Gsixverb\% on Google, and under the strict and verb-only variants the two corpora lie within \CsGapMax points of each other; per distinct pair all three variants lie within \CsGapMaxD points. In \CsAbove of the \CsCells Graph cells the ladder collapses more
pairs than the vendor level does, the exception being the write-heavy variant
on \textsf{Application} (\CsAppWh\% against \MAppCollapsed\%). Pooled over tasks, the write-heavy variant collapses fewer Graph pairs (\CsDWWh\%, \CsAppWh\%) than Google pairs (\Gsixwriteheavy\%). The difference between the Graph and Google rows is therefore largely a property of the scales: Microsoft's level separates more incomparable pairs than a ladder
computed from verbs and names. The corpora do differ on the baseline, where the
ladder rates \CsGraphBaseLo--\CsGraphBaseHi\% of Graph's comparable pairs equal
against \GsixBaseLo--\GsixBaseHi\% of Google's, so comparable Graph pairs are
easier for any of these scales to separate.

The Graph collapses include pairs whose reach differs thirtyfold. Under \textsf{Application}, the task of listing and creating OAuth2 permission grants is minimally authorised either by
\texttt{Directory.ReadWrite.All}, reaching 355 endpoints, or by
\texttt{DelegatedPermission\allowbreak Grant.\allowbreak ReadWrite.All}, reaching 11. Both are
\texttt{privilegeLevel} 4 (Figure~\ref{fig:collapse}). Under \textsf{DelegatedWork}, the
permissions \texttt{DeviceManagement\allowbreak ManagedDevices.\allowbreak ReadWrite.All} (598
endpoints) and \texttt{DeviceManagement\allowbreak Configuration.\allowbreak ReadWrite.All}
(2{,}177 endpoints) are likewise both level 3.

\begin{figure}[t]\centering
\includegraphics{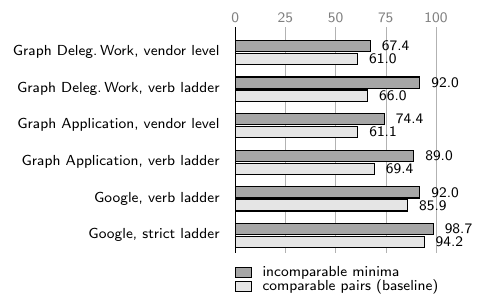}
\caption{Share of grant pairs rated equal over incomparable minima (dark) and
comparable pairs (light), per pair; the values are cells of
Table~\ref{tab:scales}. On the same Graph pairs the verb ladder collapses more
than Microsoft's level and about as much as on Google.}
\label{fig:rates}
\end{figure}

Microsoft's second risk signal adds information the level lacks: on \MDWAdmin \textsf{DelegatedWork} pairs the competing minima differ in whether they require administrator consent, and on \MDWAdminSame of them the level gives both grants the same value.

\section{Measure Dependence of the Ranking}
\label{sec:measures}

Where a scale orders an incomparable pair, we test whether the order survives a change of measure, scoring each grant under four risk measures. On Google, \emph{breadth} is the
number of methods authorised; \emph{write} is how many of those change state,
read from the HTTP verb with read-style custom methods excluded, since Google's
API conventions give a \texttt{POST} to read-only calls that need a request
body; \emph{services} is how many distinct APIs are reachable; and
\emph{admin} is whether any administrative surface is reached. On Graph the
operation is the endpoint: breadth counts endpoints, write counts non-\texttt{GET}
endpoints with read-style \texttt{POST} actions such as
\texttt{getMemberGroups} excluded, services counts distinct top-level path
segments (\texttt{users}, \texttt{groups}, \texttt{deviceManagement} and so
on), and admin applies the same name pattern to endpoint paths.

Of \GscalPairs incomparable minima pairs on Google, \GscalConf
(\GscalConfPct\%) are ordered in opposite directions by two of these measures.
On Graph the rate is \MsDWConf\% of \MsDWN \textsf{DelegatedWork} pairs and
\MsAppConf\% of \MsAppN \textsf{Application} pairs, and none of the \MsDPN
\textsf{DelegatedPersonal} pairs. Counted once per distinct pair the rates are
\GscalConfD\% of \GscalDistinct on Google and \MsDWConfD\% and \MsAppConfD\% of
\MsDWDN and \MsAppDN on Graph. Every Google pair, and all but \MsDWUnsep pair in
each well-powered Graph mode, is separated by at least one of the four
measures. On such pairs the ranking depends on the measure, which a single scale value does not record.

The conflict does not depend on the write heuristic: counting every
non-\texttt{GET} verb as a write instead moves the Google figure to
\GscalConfRaw\%, and on Graph, counting writes from the raw verb or reading
admin from the vendor's \texttt{requiresAdminConsent} flag keeps it within
\MsRobLo--\MsRobHi\%. It also does not hinge on one measure: taking the measures two at a time on Google, \GscalSubNZ of the \GscalSubTot pairings disagree on some
pair, at rates from \GscalSubLo\% to \GscalSubHi\%, and dropping
\emph{services}, the measure that contributes most, still leaves breadth and
write disagreeing on \GscalNoServices\% of pairs (\MsDWBW\% and \MsAppBW\% on
Graph). The most frequent disagreements on Google are write against services
(\GscalTopA pairs), breadth against services (\GscalTopB) and breadth against
write (\GscalTopC).

The BigQuery pair of \S\ref{sec:antichain} is the canonical Google instance. By breadth, \texttt{bigquery} is smaller, \ExBqBreadth against \ExCpBreadth. By state-changing count, \texttt{cloud-platform.read-only} is smaller, \ExCpWrite against \ExBqWrite. On Graph,
managing application registrations under \textsf{DelegatedWork} is minimally
authorised by \texttt{Application.ReadWrite.All}, reaching \MsExBreadthA
endpoints of which \MsExWriteA change state, or by
\texttt{AppRegistration.Create} with \texttt{Directory.Read.All}, reaching
\MsExBreadthB of which \MsExWriteB change state. 

\section{Sources of Incomparability}
\label{sec:sources}

\begin{table}[t]\centering\small
\begin{tabular}{@{}lrrrr@{}}
\toprule
Corpus and mode & pairs & breadth & domain & granul. \\
\midrule
\multicolumn{5}{@{}l}{\emph{per pair, pooled over tasks}}\\
Google, resources  & \SrcGN    & \SrcGBe\%    & \SrcGDc\%    & \SrcGGr\% \\
Graph, Deleg.Work  & \SrcMDWN  & \SrcMDWBe\%  & \SrcMDWDc\%  & \SrcMDWGr\% \\
Graph, Application & \SrcMAppN & \SrcMAppBe\% & \SrcMAppDc\% & \SrcMAppGr\% \\
\midrule
\multicolumn{5}{@{}l}{\emph{per distinct pair}}\\
Google, resources  & \SrcGND    & \SrcGBeD\%    & \SrcGDcD\%    & \SrcGGrD\% \\
Graph, Deleg.Work  & \SrcMDWND  & \SrcMDWBeD\%  & \SrcMDWDcD\%  & \SrcMDWGrD\% \\
Graph, Application & \SrcMAppND & \SrcMAppBeD\% & \SrcMAppDcD\% & \SrcMAppGrD\% \\
\bottomrule
\end{tabular}
\caption{What makes two minimal sufficient grants incomparable, classified
mechanically in priority order: breadth versus effect, domain crossing,
granularity. Granularity is the residual class, since once the first two tests
fail minimal footprints cannot nest, so the split describes composition rather
than coverage.}
\label{tab:sources}
\end{table}

To locate its sources, we classify each incomparable pair mechanically, in priority order:
\emph{breadth versus effect}, where one grant reaches more operations while the
other changes more state; \emph{domain crossing}, where the sets of services
reached are mutually non-containing; and \emph{granularity}, where one grant's
service set contains the other's but its operation set does not.

Table~\ref{tab:sources} reports the split. Granularity is the residual class by
construction, so its share is what the first two tests leave behind rather than
an independent finding; the informative quantities are the first two columns.
Counting each pair once
moves the shares by up to \SrcShiftMax points and leaves granularity and domain
crossing as the two large classes in every corpus.

Granularity accounts for \SrcGrLo--\SrcGrHi\% of pooled pairs. The recurring
shape is a coarse permission over a narrow area competing with a fine
permission over a wide one, and the OAuth2 grants task of \S\ref{sec:scales} is
an instance: \texttt{Directory.ReadWrite.All} is coarse and central,
\texttt{DelegatedPermission\allowbreak Grant.\allowbreak ReadWrite.All} is fine and peripheral, and
each is minimal.

In principle every antichain is removable by the vendor: for incomparable
sufficient footprints $A$ and $B$ the task lies in $A \cap B$, so a permission
with that footprint would sit strictly below both. The classes differ in
whether such a permission is a product a vendor would ship. For granularity it
is an intermediate scope over an existing area. For domain crossing, at
\SrcDcLo--\SrcDcHi\%, it spans parts of the estate the catalogue keeps apart,
and for breadth versus effect, at \SrcBeLo--\SrcBeHi\%, it is a credential at
once narrow in reach and limited in what it can change, a shape few catalogues
offer.

\section{Undetermined Reach Under a Perfect Score}
\label{sec:perfect}

Take an agent that always selects a minimal sufficient grant. It never
over-privileges in the sense of footprint minimality, and since the tested
scales rate most incomparable minima equal, most of its selections are also
indistinguishable to a benchmark built on them. On an antichain it still has a
choice, and this section measures what that choice is worth, in operations
authorised, per distinct antichain (\S\ref{sec:method}). On Google, the \GrN antichains behind \GrTaskN tasks give a median
ratio of \GrMed{}$\times$; per task the median would read \GrTaskMed{}$\times$,
because one wide antichain sits under a deep resource tree. On Graph the two
units agree closely (\MDWrMed{}$\times$ per antichain and \MDWrTaskMed{}$\times$
per task under \textsf{DelegatedWork}).

Catalogue design moves the rate of antichains and their spread together.
Google's fine-grained scopes give fewer and narrower antichains, with
\GrGeTen\% above tenfold between widest and narrowest minimum. Graph's denser
overlap gives wider spreads, median \MDWrMed{}$\times$ and \MApprMed{}$\times$ over
\MDWrN and \MApprN distinct antichains, with \MDWrGeTen\% above tenfold under
\textsf{DelegatedWork}.

The extremes occur in widely used services. Editing a Microsoft Teams channel message
is minimally authorised by grants reaching 6 and 265 operations respectively, a
factor of \MDWrMax. On Google, managing Firebase projects is minimally authorised either by a grant reaching 248 operations or by one reaching
12{,}247.

A perfect minimality score is therefore compatible with a blast radius that
varies tenfold or more on \GrGeTen--\MDWrGeTen\% of distinct antichains in the Google resource family and the two main Graph modes, and the score does not record which end of that range the agent reached.

\section{Per-Step Versus Whole-Workflow Minima}
\label{sec:composition}

The MCP authorisation guidance recommends acquiring scopes progressively
\cite{mcpauth}, and per-call least-privilege solvers make the analogous
per-step choice \cite{miniscope}. A natural reading of the results above places
the difficulty in authority accumulating step by step.

On \GwfN synthetic workflows of $k \in \{2,3,4,6,8\}$ methods drawn within one
of \GwfApis Google APIs, a selector that greedily takes the narrowest minimum
available at each step ends holding the whole-workflow minimum on
\GwfSuccLo--\GwfSuccHi\% of workflows, with median excess zero. The adversarial
selector, taking the widest minimum at each step, exceeds it on
\GwfWorstPct\%. These two selectors bracket the range an implementation is likely to sit
in.

Google's heavily nested, fine-grained scopes absorb the accumulation that the
structure permits in principle, which leaves the single-step choice as the
source of the spread.

\section{A Reporting Format for the Choice}
\label{sec:format}

A benchmark already enumerates the antichain, since that is how it knows the
agent stayed minimal, so it can report two things instead of one. The first is
the existing minimality rate, a property of the agent's restraint that today's
benchmarks measure well. The second is which member of the antichain the agent
selected, reported as a vector over the four risk measures of \S\ref{sec:measures}. A single coordinate would not do: breadth, state-changing count, services reached and
administrative surface rank the same pair in opposite directions on
\ConfAllLo--\ConfAllHi\% of pairs, so any normalisation into one coordinate reintroduces an unreported tie-break.

\paragraph{Experimental design.}
To see whether the choice is worth recording we put \AgModels frontier models,
from as many vendors, choosing among grants enumerated from the tables. Each decision is a
closed choice among enumerated grants, whereas a deployed agent composes a
scope request. The closed format is deliberate: in a free-form request a model
can also fail by requesting too little or by naming a scope that does not
exist, and a score would mix those failures with the choice among minimal
grants. Enumerating sufficient grants isolates the decision this paper
concerns, which of several sufficient grants to hold. The results describe what
a closed benchmark can observe about that decision, not how often free-running
agents over-request.

\paragraph{Choice within an antichain.}
Each model was shown a task and the whole antichain of minimal sufficient
grants for it, and asked to pick one: \AgN\ decisions over \AgAnti distinct
antichains from both corpora, at temperature zero, with option order shuffled
per antichain and no failed calls. Under the \textsf{names} condition a model
sees only the permission names, which is what a deployed agent has when it
requests a scope. Under \textsf{reach} it is additionally told how many
operations each grant authorises.

Under \textsf{names} the models agree unanimously on \AgNmUnan\% of antichains,
with mean pairwise agreement \AgNmPair\%. They select the narrowest available
grant \AgNmNarrow\% of the time and the widest \AgNmWide\%, and \AgNmGeTwo\% of
selections authorise at least twice what the narrowest option would have.
Every one of these selections is minimal, so every one of them scores perfectly
on the benchmarks of \S\ref{sec:background}.

Showing reach changes picks more than it changes agreement. Choice of the narrowest rises by \AgNarrowGain\ points to
\AgRcNarrow\%, and the pick changes on \AgMoved\% of model-antichain pairs. Unanimity rises by \AgUnanGain\ points, to
\AgRcUnan\%, and the models split: \AgRcTopN of the \AgModels now take the narrowest
on \AgRcTopLo--\AgRcPmHi\% of decisions while the other \AgRcRestN reach \AgRcPmLo--\AgRcRestHi\%.

\paragraph{Mixed pools of minimal and larger grants.}
That experiment tells the model implicitly that every option is minimal, so no larger grant has to be ruled out. We ran it again with the pool mixed:
\MxTasks Graph tasks, each offered its minimal sufficient grants together with
grants that are sufficient but strictly larger, unlabelled and shuffled, and
the instruction to take the one granting least additional authority, for \MxN\
decisions. The models select a minimal grant on \MxMinimal\% of decisions,
\MxMinLo--\MxMinHi\% across the six, close to a perfect score on the question
over-privilege benchmarks ask. They select the narrowest grant in the pool on
\MxNarrow\%. Everything in the \MxGap-point gap between those rates is minimal
and correct by the benchmark's own definition, so a minimality score cannot
express it. Per model the gap runs \MxGapLo--\MxGapHi\ points, the widest
belonging to a model minimal on \MxWorstMin\% of decisions and narrowest on
\MxWorstNarrow\%. Across all six, \MxGeTwo\% of selections authorise at least
twice what the narrowest option in the same pool would have.

The choice also varies within a model. Resampling \ScK times at temperature one
over \ScN further decisions, a model returns the same member of the same antichain every time in \ScAll\% of the \ScCells model-antichain cells, from \ScLo\% to \ScHi\% across the six.

\begin{table}[t]\centering\small
\setlength{\tabcolsep}{3.2pt}
\begin{tabular}{@{}lrrrrrr@{}}
\toprule
Condition ($n$) & bread. & write & serv. & admin & all 4 & dom. \\
\midrule
names (\FmNmN) & \FmNmBreadth & \FmNmWrite & \FmNmSvc & \FmNmAdmin & \FmNmAllFour & \FmNmDom \\
reach (\FmRcN) & \FmRcBreadth & \FmRcWrite & \FmRcSvc & \FmRcAdmin & \FmRcAllFour & \FmRcDom \\
mixed (\FmMxN)       & \FmMxBreadth & \FmMxWrite & \FmMxSvc & \FmMxAdmin & \FmMxAllFour & \FmMxDom \\
\bottomrule
\end{tabular}
\caption{Model choices scored under each risk measure of \S\ref{sec:measures}
(\%): antichain-only pools under \textsf{names} and \textsf{reach}, and mixed
pools. The first four columns give the share of decisions that pick the
smallest available grant under that measure, ties counting as smallest;
\emph{all 4} under every measure at once; \emph{dom.}\ the share picking a
Pareto-dominated grant, one that another option matches on every measure and
beats on at least one. Re-scored from stored replies, with no new model
calls.}
\label{tab:fourmeasures}
\end{table}

\paragraph{The choices under all four measures.}
The rates above call a pick narrowest by breadth alone. Table~\ref{tab:fourmeasures}
re-scores the same stored replies under each of the four measures, with option
footprints recomputed from the vendor tables and checked against the counts
the \textsf{reach} prompts displayed (\FmMismatch mismatches in \FmChecked). The
other measures give the same picture as breadth. Under \textsf{names} the
models pick the smallest grant by state-changing count on \FmNmWrite\% of
decisions and by services reached on \FmNmSvc\%; admin is a binary measure and
usually tied. A grant that is smallest under all four measures at once is on
offer in \FmNmAllFourEx\% of these decisions and taken in \FmNmAllFourGivenEx\%
of those. On \FmNmDom\% of decisions under \textsf{names} and \FmMxDom\% in the
mixed pool, the model takes a grant that another option matches or beats on
every measure, a choice that no positive weighting of the four measures prefers; nearly all such picks are minimal and score perfectly. In the mixed pool the gap
between minimality and the grant smallest under all four measures is
\FmMxGapAllFour\ points, wider than the \FmMxGapBreadth-point gap under breadth
alone.

Where the antichain is small, listing it outright is more informative than any
summary, and it gives a deployment reviewer each grant that would have worked together with its footprint.

\section{Related Work}
\label{sec:related}

\paragraph{Over-privilege benchmarks.} The systems whose construction we test
are the direct context \cite{toolprivbench,fortis,authbench}, alongside
enforcement work that detects unnecessary privileged actions rather than
scoring a choice among sufficient ones \cite{skillscope}. We do not dispute their labels; the uniqueness they construct holds in their inventories but not in vendor tables, so metrics built on it inherit an unreported tie-break. AuthBench is closest to
our position, defining a genuine partial order over read, write and execute
and observing that ``the task-sufficient boundary is therefore not a fixed
ground truth determined by the task instruction alone'' \cite{authbench}. Its
non-uniqueness has a different source, namely different agents solving a task
by different routes; ours holds for a fixed task and a fixed route, and is a
property of the authorisation table. Closer still, \citet{earnedauthority}
state in passing that a task's authority ``need not have a unique minimum
because alternative plans may require incomparable resources or effect types''.
They reach it from the evolution of an agent's entitlements over time and do
not measure it; the antichain here is visible in a static table, in one step.

\paragraph{Composition safety.} ChainCaps prevents authority from being
amplified across a tool chain by propagating sink budgets under intersection,
and proves a non-amplification theorem \cite{chaincaps}. APPA addresses the
availability cost of monotone taint tracking, which ``either over-blocks benign
operations or permanently strands downstream execution'' \cite{appa}. Both
concern an upper bound on authority, whereas the antichain is a statement about
minimality; a system can be non-amplifying and still face our selection
problem, since the choice of which minimal grant to start from is prior to any
propagation rule.

Label creep in coarse-grained information-flow systems, where a program
becomes ``too tainted to do anything useful'' \cite{vassena2018}, concerns one
execution drifting over time and can occur with a unique sufficient authority.

\paragraph{Classical access control.} Least privilege was stated informally
\cite{saltzer1975}, and a formal treatment argues that the minimum is not
generally computable \cite{schneider2003}; neither supplies a uniqueness
condition. Lattice models concern the well-definedness of \emph{labels}
\cite{denning1976}, and decentralised labels form a preorder rather than a
lattice \cite{myers2000}; neither asks whether the authorities sufficient for a
task are closed under meet. The nearest structural precedent is role
mining, where minimising a role set is NP-hard \cite{vaidya2007} and the two
natural objectives yield a Pareto front of mutually non-dominated solutions
rather than one optimum \cite{crampton2024}, though the object minimised there
is an RBAC decomposition rather than the authority for a task.

\section{Implications for Practice}
\label{sec:implications}

\paragraph{Separate reporting of restraint and tie-break.} A benchmark that
scores over-privilege on a total scale reports a composite of whether the agent
stayed within the minimal sufficient authorities and which member of the
antichain it picked. The second is a property of the tie-break, which on our
corpora the scale leaves undecided on \MCollapsedLo--\MCollapsedHi\% of Graph
pairs under the vendor level and \CsGoogleLo--\CsGoogleHi\% of Google pairs
under the ladder.

\paragraph{The tie-break as declared policy.} Choosing among incomparable
minima is a risk-preference decision, such as narrower reach against fewer
state changes. Systems that compute a minimal permission set should show the
alternatives to whoever approves the grant; MiniScope already asks a user to
approve the set it computes \cite{miniscope}, and showing the incomparable
alternatives at that prompt would let the accountable approver make the choice.

\paragraph{Catalogue refinement as a vendor lever.} Because \SrcGrLo--\SrcGrHi\% of incomparable pairs fall in the granularity class (\S\ref{sec:sources}), a vendor holds a lever the benchmark designer does not.
Adding an intermediate permission between a coarse central grant and a fine
peripheral one removes those antichains at the source. The table checks
need no annotation and no model calls, so a vendor can run them on its own
catalogue to find where its ordinal level fails to separate grants that differ
in reach.

\section{Limitations}
\label{sec:limits}

The coverage order is derived from method-level tables and does not see
resource-level or conditional constraints; since finer granularity can only add
incomparability, the widths reported here are lower bounds.

Tasks are generated from vendor structure, namely resource method sets, random
$k$-subsets within an API and Graph path groups, not observed from deployed
agent traces. The rates are properties of the tables under these families, and
a workload with a different distribution over operations would give different
rates; the mechanism, a task authorised by several mutually non-containing
grants, does not depend on that distribution.

The ladder of \S\ref{sec:scales} is our mechanical reconstruction of FORTIS's
stated semantics, not FORTIS's own labels, which do not exist for these APIs;
applying it to both corpora controls for the scale in the comparison without
making it FORTIS's. It aggregates by maximum, the ordinal coordinate of
FORTIS's objective, so Table~\ref{tab:scales} does not audit the cardinality
tie-break after it; the flagship collapse involves single permissions and is
unaffected by aggregation. On Graph the four risk measures of
\S\ref{sec:measures} are analogues of the Google ones: services counts
top-level path segments, which group endpoints by resource family rather than
by product.

The Wilson interval on the \textsf{DelegatedWork} non-uniqueness rate is
\CIDWNonuniqLo--\CIDWNonuniqHi\%, and \textsf{Application} is comparably tight,
but \textsf{DelegatedPersonal} rests on \MDPTasks tasks and \MDPPairs pairs, so its rows are
indicative only. Section~\ref{sec:composition} uses Google alone, since the
Graph file lacks the per-step structure, and describes Google's scope design
rather than per-step acquisition in general.

A third corpus delimits the result: 1{,}500 tasks across
1{,}263 APIs.guru specifications gave no non-unique task, because OpenAPI's
\texttt{security} field mostly encodes the authentication mechanism rather than
the authority carried, so its alternatives are equal rather than incomparable.

The choice experiments measure the ranking decision rather than the request
path (\S\ref{sec:format}). The mixed pool's non-minimal options come from the
hitting-set enumeration or, where that supplied too few, from extending one
minimal grant by one permission, a structure a real over-request need not have.
Headline rates are single samples at temperature zero; self-consistency comes
from a temperature-one resample over a random half of the antichains.

\paragraph{Data and disclosure.} Both corpora are authorisation tables their
vendors publish for developers to read, and nothing in this paper describes a
vulnerability, a bypass, or an exploitable configuration. The findings concern
the resolution of a published ordinal scale, not the correctness of any access
decision. The released package fixes the retrieval date, the upstream commit
for the Microsoft file, per-document hashes for the Google corpus, the seed,
and the model identifiers, sampling settings and raw replies of
\S\ref{sec:format}. Large language models were used to draft and revise prose and to
write analysis scripts; every number, script and claim was checked by the
author against the released artefacts.

\section{Conclusion}

Least privilege for an agent is usually posed as a minimisation with a unique
answer. On the tables Google and Microsoft publish, it is not unique for \MNonuniqLo--\MNonuniqHi\% of Microsoft Graph tasks
and \GnonuniqRPct\% of Google ones, and the competing answers always authorise
different operations. The ordinal scales that agent evaluation uses to force
uniqueness mostly respond by giving those answers the same level, and applying
one scale to both corpora shows that how often they do so depends on the scale more than on the corpus; Microsoft's own scale gives one level to grants reaching 355 and 11 endpoints. Which of the minimal sufficient authorities an
agent holds is a security-relevant decision without a canonical rule, and we
recommend that benchmarks report it as a vector beside minimality and that
deployments show the alternatives to whoever approves the grant.

\bibliography{refs}

\end{document}